%% file: glowinski_arxiv.tex
\documentclass[12pt,reqno,twoside]{amsart}

\usepackage{amsmath,amsthm,bm}
\usepackage{hyperref}
\usepackage[margin=1in]{geometry}

\input abbrevs
\input epsf

\newtheorem{theorem}{Theorem}
\newtheorem{lemma}[theorem]{Lemma}

\usepackage[textsize=small]{todonotes}

\begin{document}

\title[Adjoint-Based Estimation of Sensitivity]{Adjoint-Based Estimation of Sensitivity of Clinical Measures to Boundary Conditions for Arteries}

\thanks{Dedicated to Prof. Roland Glowinski. \\ This work is partially supported by NSF grant DMS-2110263 and the AirForce Office of Scientific Research under Award NO: FA9550-22-1-0248.}

\author{Rainald L\"ohner, Harbir Antil, Juan Cebral, Fernando Mut}
\address{R. L\"ohner and H. Antil. 
Center for Computational Fluid Dynamics and 
Center for Mathematics and Artificial Intelligence,
4400 University Dr., 
George Mason University,
Fairfax, VA 22030-4444, USA}
\address{J. Cebral and F. Mut. 
Dept. of Biomedical Engineering, George Mason University, 
4400 University Dr., 
George Mason University,
Fairfax, VA 22030-4444, USA
}

\begin{abstract}
The use of adjoint solvers is considered in order to obtain the sensitivity of
clinical measures in aneurysms to incomplete (or unknown) boundary conditions
and/or geometry. It is shown that these techniques offer
interesting theoretical insights and viable computational tools to obtain
these sensitivities.
\end{abstract}


\maketitle

\section{Introduction}
\begin{figure}
    \centering
    \includegraphics[width=10cm]{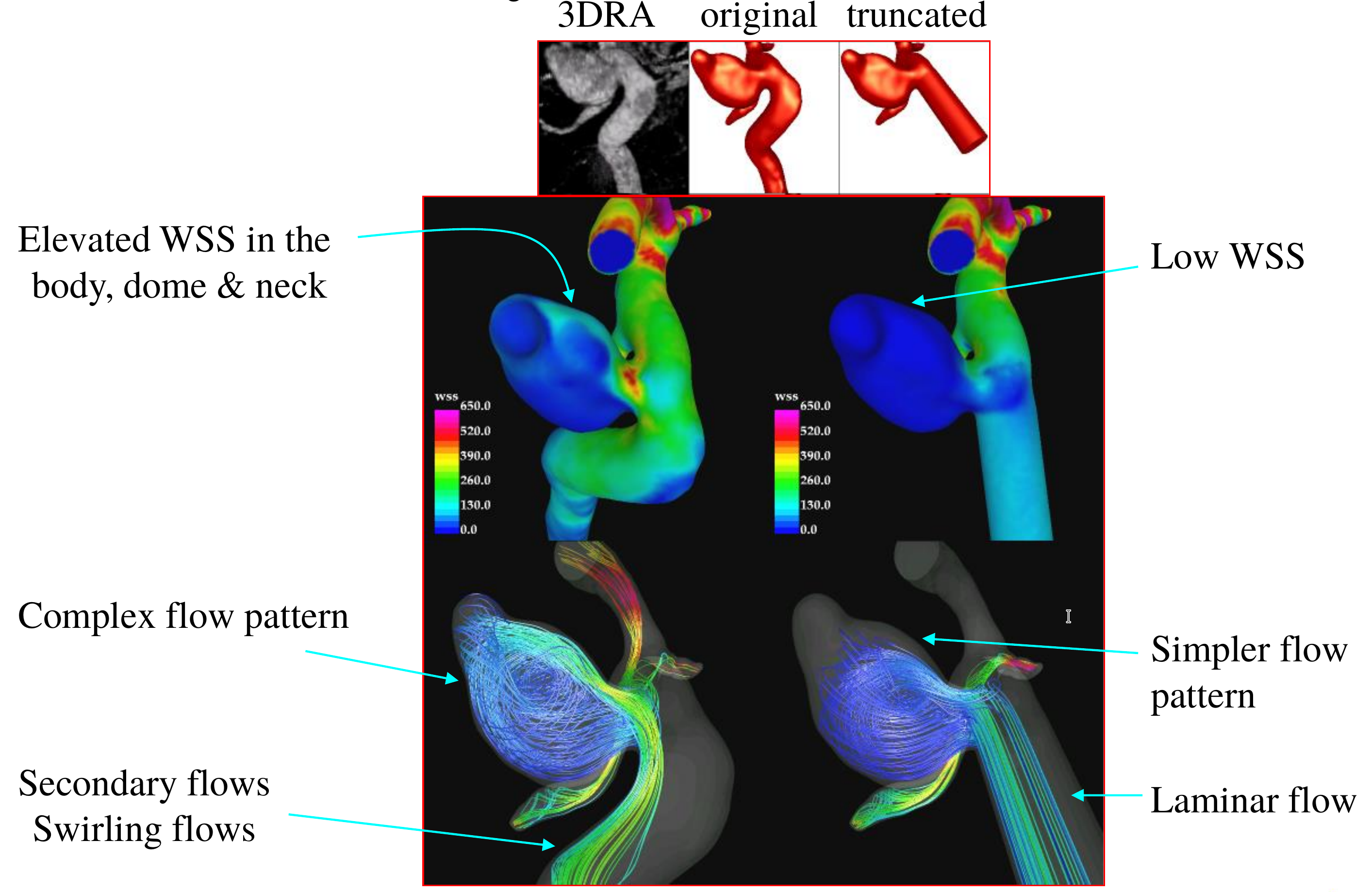}
    \caption{Vessel 1: Difference in flow features between properly resolved and unresolved upstream geometry.}
    \label{fig:1}
\end{figure}

\begin{figure}
    \centering
    \includegraphics[width=10cm]{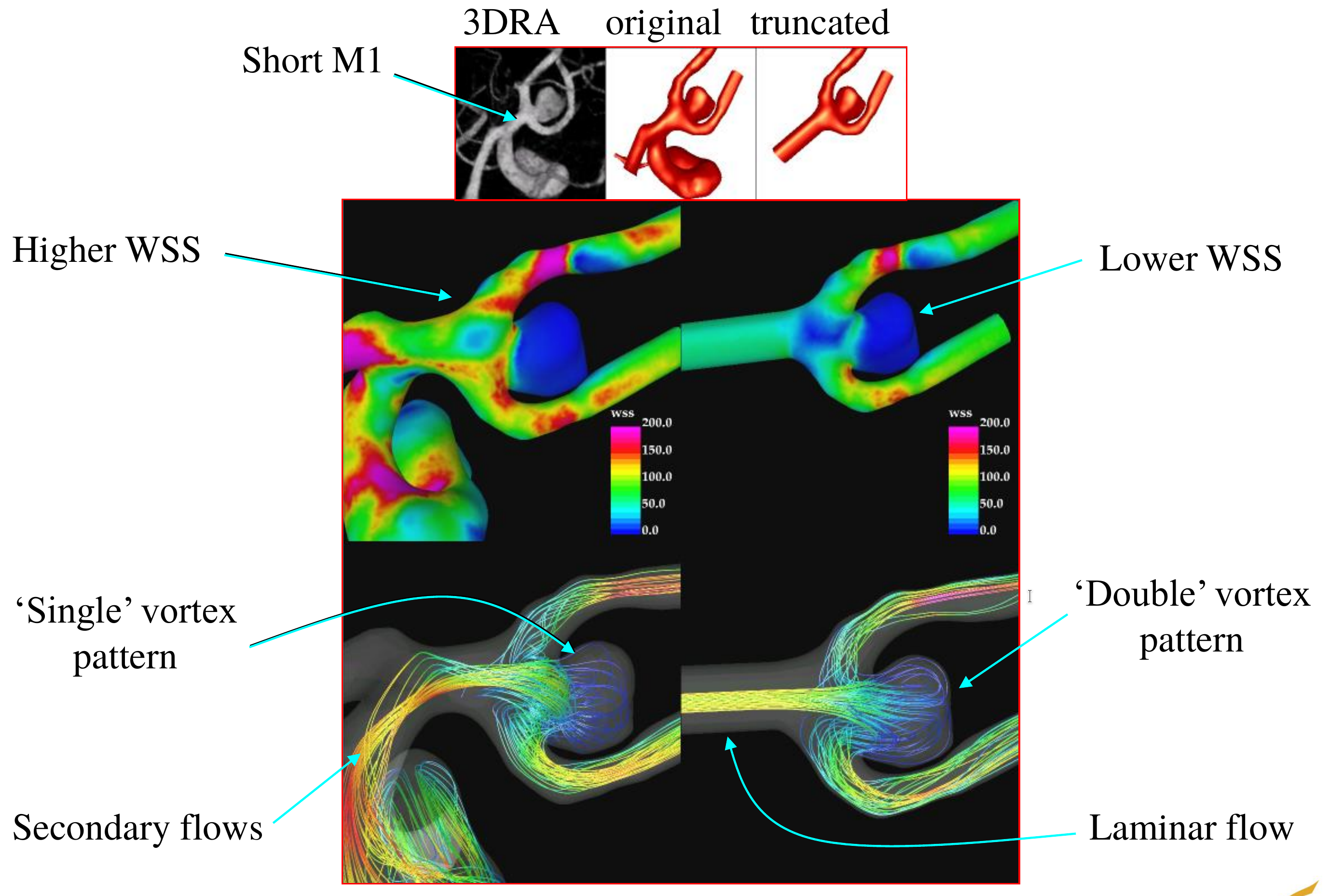}
    \caption{Vessel 2: Difference in flow features between properly resolved and unresolved upstream geometry}
    \label{fig:2}
\end{figure}

The analysis of haemodynamic phenomena and their clinical relevance
via computational mechanics (fluids, solids, $\dots$) is now common
in research and development. Yet a recurring question has been
the influence of boundary conditions and geometry on
`clinically relevant measures'. As an example, consider flows in
aneurysms. A crucial question is how far upstream the geometry has
to be modeled accurately in order to obtain sufficiently accurate
flow predictions, as well as their associated loads on vessel walls
(shear, pressures) and clinically relevant measures (such as kinetic and
vortical energy, vortex line length, etc.). In many cases,
users may not have sufficient upstream information, so this question
is of high relevance. The thesis of Castro and subsequent publications
\cite{castro2006computational,cebral2003blood} have shown how dramatic the difference between well
resolved
upstream geometries and so-called `cut' geometries can be. In some cases,
completely different types of flow were seen, which in turn could have
led to different clinical decisions. Figures~\ref{fig:1}-\ref{fig:2} show two examples.

To complicate matters further, the flow is transient/pulsating,
and the flowrate and flow profile coming in at the upstream boundary
in most cases is
unknown. It is a common practice to simply set some kind of pipe
flow profile (Poiseuille, Womersley) at the inflow,
adjusting the analytical parameters to the estimated/known flux. \\
The central question remains: \emph{what is the influence of a change of
boundary conditions (e.g. inflow profiles) or geometry (e.g. more
upstream/downstream geometry) on the clinically relevant measures~?} \\
A simple way to answer this question is to perform several runs, each
with a different geometry or different boundary condition. This finite
difference approach can then yield the sensitivity of a `measure of
clinical relevance' $I$  to a change in geometry or boundary
condition $\bm z$. Another possibility is via adjoints
\cite{JLLions_1971a,FTroeltzsch_2010a,KIto_KKunisch_2008a,MHinze_RPinnau_MUlbrich_SUlbrich_2009a,
HAntil_DPKouri_MDLacasse_DRidzal_2018a}.
We also refer to a series of works by Glowinski and collaborators on the 
role of adjoints in optimization \cite{MR1646758,MR1803930,MR1892232,HAntil_RGlowinski_RHWHoppe_CLinsenmann_TWPan_AWixforth_2010a,MR4238769,MR4416986}.
See also \cite{MR1460653,MR1638072,MFGunzburger_LSHou_TPSvobody_2008}. We emphasize that this list is incomplete
as many authors have made fundamental contributions to this topic.

\subsection{Upstream Boundary Conditions for the Flow}
It is known from empirical evidence and simple fluid mechanics that
given any steady inflow velocity profile, after a given number of
diameters along the pipe the flow will revert to a simple pipe flow
(Poiseuille). This so-called hydrodynamic entry length $L_h$ is a
function of the Reynolds number $Re$, and for laminar flow and
uniform inflow is given by:
\begin{equation}
L_{h} = 0.05 Re \, D \, , \quad Re = \frac{\rho U_e D}{\mu} \, ,
\end{equation}
where $\rho, U_e, \mu$ denote the density, mean entrance velocity
and viscosity of the flow and $D$ the vessel diameter.
For blood and a typical artery $\rho=1~g/cm^3, U_e=50~cm/sec, 
\mu=0.04~g/cm/sec,
D=0.1~cm$, so $Re=O(100)$ and $L_{h} = 5~D$. Note that this estimate
is only valid for steady flows and a uniform inflow. As far as the
authors are aware, similar estimates for vessels with high
curvatures (tortuosity) as typically encountered in arteries are not
available.
We note in passing that for the unsteady cases analyzed by  \cite{castro2006computational,cebral2003blood} the number of 
upstream diameters required 
before the flow did not change in the aneurysms was much higher 
than the estimate given above.

\subsection{Possible Mathematical Approaches}
In order to formulate the problem mathematically, we can consider
different approaches.
\begin{itemize}
\item[a)] \ub{Empirical Data}: for any given geometry/case, one could
perform a series of studies, changing the type of
inflow (vortical flows, unsteady flows) and seeing how long the
observed hydrodynamic entry lengths are;
\item[b)] \ub{Sensitivity Analysis I}: one could try to obtain a
`topological derivative' that
measures the sensitivity of the flow in the aneurysm with respect
to movement of the upstream boundary.
\item[c)] \ub{Sensitivity Analysis II}: one could obtain a
`flow derivative' that measures the sensitivity of the clinical measure
of the flow in the aneurysm with respect
to changes of the entry flow in the upstream boundary.
\end{itemize}

\medskip
\noindent
{\bf Outline:} The remainder of the paper is organized as follows. In 
Section~\ref{s:general}, we first introduce a generic optimization problem 
formulation and adjoint framework. This generic discussion is well-known. 
This is followed by an example of Navier-Stokes specific to the 
aneurysm problem. We study the sensitivity with respect to the inflow
velocity and inflow position. Section~\ref{s:numimp} focuses on numerical 
implementation. In Section~\ref{s:Poiseuille}, we present a specific
example corresponding to the 2-D channel flow. For this example, we
are able to derive explicit expressions for the state variables, adjoint 
variables, and the sensitivities (see Appendix~\ref{app:analytical}). This 
is followed by a realistic aneurysm example in Section~\ref{s:aneu}, where 
we study the sensitivity of the `measure of clinical relevance' $I$. All 
the numerical examples confirm the proposed approach.

\section{General Adjoint Formulation}
\label{s:general}

Suppose we have a `measure of clinical relevance' $I$ for a region
that is in or close to an aneurysm. This could be the kinetic or
vortical energy, the shear stress or the length of vortex lines -
all of which have been proposed in the literature 
\cite{mut2011computational,detmer2019associations,detmer2020incorporating}.

The question then becomes:
how sensitive is this measure to the (often unknown) boundary
conditions imposed or the (often approximate) geometric accuracy~?
Given that $I$ is a function of the unknowns $u$ and these
in turn are a function of a set of parameters $z$ describing
the boundary conditions or the geometry, the answer to this
question is given by the gradient of $I$. Consider the  
well-known generic minimization problem 
\[
    \min_{u,z} I(u,z) \quad \mbox{subject to} \quad
    e(u,z) = 0 \, ,
\]
where $I : U \times Z \rightarrow \mathbb{R}$ is the cost functional
and $e(\cdot,\cdot) : U \times Z \rightarrow Y$ is the PDE constraint.
Here $U,Y$ and $Z$ are function spaces. Typically, $U,Y$ are Banach spaces 
and $Z$ is a Hilbert space. Under very 
generic conditions, one can establish existence of solution to the
above optimization problems, see 
\cite{MHinze_RPinnau_MUlbrich_SUlbrich_2009a,HAntil_DPKouri_MDLacasse_DRidzal_2018a}.
As it has been known in the literature, there are two ways to derive 
the expression of the adjoint and the gradient of objective function 
$I$. The first approach is the so-called reduced formulation, where 
assuming that the PDE is uniquely solvable, one considers the 
well-defined control-to-state map 
\[
    z \mapsto u(z) 
\]
with $(u(z),z)$ solving the PDE $e(u(z),z) = 0$. The reduced objective
functional is then given by $\mathcal{I}(z) = I(u(z),z)$. Then one 
obtains the derivative of $\mathcal{I}$ with respect to $z$ which 
also requires computing the sensitivites of $u$ with respect to $z$.
The second approach is the full space formulation and it 
requires forming the Lagrangian. Under fairly generic conditions
(constraint qualificiations), one can establish the existence of
Lagrange multipliers in this setting, see 
\cite{JZowe_SKurcyusz_1979a,MHinze_RPinnau_MUlbrich_SUlbrich_2009a}. Regardless, in both cases, the same expression of gradient is obtained \cite[Pg.~14]{HAntil_DPKouri_MDLacasse_DRidzal_2018a}.

We briefly sketch the Lagrangian 
approach and refer to \cite{MHinze_RPinnau_MUlbrich_SUlbrich_2009a,HAntil_DPKouri_MDLacasse_DRidzal_2018a} 
for details. Let $p$ denotes the adjoint variable, then the Lagrangian functional is given by 
\begin{equation}\label{eq:Lag}
    L(u,z,p) = I(u,z) - \langle e(u,z) , p  \rangle_{Y,Y^*} \, .
\end{equation}
Then at a stationary point $(u,z,p)$ the following conditions hold
\begin{equation}\label{eq:Lagopt}
\begin{aligned}
    L_p(u,z,p) &= 0 ,  \\
    L_u(u,z,p) &= 0 ,  \\
    L_z(u,z,p) &= 0 . 
\end{aligned}    
\end{equation}
Our goal for the application under consideration is not to solve the 
above optimization problem, but rather
derive the expression of the gradient $L_z(u,z,p)$. In view of the 
expression of the Lagrangian given in \eqref{eq:Lag}, it is 
not difficult to see that conditions in \eqref{eq:Lagopt} are 
equivalent to 
\begin{equation}\label{eq:Lagopt1}
\begin{aligned}
    e(u,z) &= 0 ,                  &&\mbox{(State equation)} \\
    e_u(u,z)^{*}p &= I_u(u,z) ,    &&\mbox{(Adjoint equation)} \\
    I_z(u,z) - e_z(u,z)^*p &= 0 .  &&\mbox{(Gradient equation)}    
\end{aligned}
\end{equation}
Namely, the gradient is given by (cf.~\cite[Pg.~14]{HAntil_DPKouri_MDLacasse_DRidzal_2018a})
\begin{equation}\label{eq:grad}
    \nabla \mathcal{I}(z) = I_z(u,z) - e_z(u,z)^*p .
\end{equation}
The consequences of the above formulation are profound:
\begin{itemize}
    \item The variation of $I$ in \eqref{eq:grad} exhibits only derivatives with respect to $z$, i.e., no explicit derivatives with respect to $u$ appear;
    \item The cost of evaluation of gradients is independent of the number of design variables (!).      
\end{itemize}

In the next section, we will apply this abstract framework to the 
case where the PDE $e(u,z) = 0$ is given by the incompressible Navier-Stokes 
equations. These equations are used to model the flow in the aneurysms.

\subsection{Incompressible Navier-Stokes and Sensitivity with Respect to Inflow}

Let the domain $\Omega \subset \mathbb{R}^d$ be sufficiently smooth,
and consisting of two subdomains $\Omega_{\rm aneurysm}$ and the remainder of the
domain $\Omega \setminus \Omega_{\rm aneurysm}$ consisting of vascular vessels. Furthermore, let the boundary $\Gamma$ of $\Omega$ consist of three parts $\Gamma_{\rm in}$ (inflow), $\Gamma_{\rm fixed }$
(fixed / wall), and $\Gamma_{\rm out}$ (outflow). Moreover, let $(\bm u, p)$ denote the velocity-pressure pair solving the incompressible Navier-Stokes equations:
	\begin{align}\label{eq:NS}
	\begin{aligned}
		 - \mbox{div}(\mu \nabla \bm u)  + (\bm{u} \cdot \nabla) \bm{u}  + \nabla p &= \bm{f} \quad  \mbox{in }  \Omega  \\
									\mbox{div } \bm{u}           &= 0 \quad  \mbox{in } \Omega \\
											 \bm{u}	       &= \bm{z}  \quad  \mbox{on }  \Gamma_{\rm in}   \\
 											\bm{u}	       &= \bm 0  \quad  \mbox{on }  \Gamma_{\rm fixed}   \\
					            (\mu \nabla \bm{u} - pI) \cdot  \bm{n} &= \bm 0  \quad  \mbox{on }  \Gamma_{\rm out}  
	\end{aligned}
	\end{align}
where $\bm f$ denotes a given force (for the current
set of applications $\bm f=0$), $\mu$ is viscosity, and $\bm n$ is the outward unit normal. Finally, $\bm z$ is some given velocity profile on the inflow boundary $\Gamma_{\rm in}$. 

Given a quantity of interest (measure of clinical relevance), 
$I(\bm u, p,\bm z)$, the goal is to obtain the derivative of $I$ 
with respect $\bm z$ with the help of adjoint 
formulation as discussed in the previous section. We begin by stating the 
following result, see \cite[Appendix~C]{MR2576248}
\begin{lemma}
\label{lem:integ}
Let $\bm u$, $\bm v$ and $\tilde{\bm u}$ be smooth vector fields, then
	\[
		\int_\Omega [ (\bm u \cdot \nabla) \bm v ] \tilde{\bm u} 
			= -\int_\Omega (\mbox{div } \bm u) (\bm v \cdot \tilde{\bm u}) 
			   + [ (\bm u \cdot \nabla) \tilde{\bm u} ] \cdot \bm v
			   + \int_{\Gamma} (\bm u \cdot \bm n) (\bm v \cdot \tilde{\bm u}) . 
	\]
When $\bm v = \bm u$ and div $\bm u = 0$, then 
	\[
		\int_\Omega [ (\bm u \cdot \nabla) \bm u ] \tilde{\bm u} 
			= -\int_\Omega  [ (\bm u \cdot \nabla) \tilde{\bm u} ] \cdot \bm u
			   + \int_{\Gamma} (\bm u \cdot \bm n) (\bm u \cdot \tilde{\bm u}) . 		
	\]	
\end{lemma}

Next, a derivation of sensitivity is provided using the adjoint approach. 
We begin by writing the Lagrangian functional 
		\[
		\begin{aligned}	
			L(\bm u, p, \tilde{\bm u}, \tilde p, \tilde{\bm u}_\Gamma) &= I(\bm u,p, \bm z) 
				- \left[ \int_\Omega \left( - \mbox{div}(\mu \nabla \bm u) 
				+ (\bm u \cdot \nabla) \bm u + \nabla p - \bm f\right) \cdot \tilde{\bm u} 
				- \tilde{p}\mbox{div } \bm u \,  {\rm dx} \right. \\
				&\quad \left.+  \int_{\Gamma_{\rm in}} (\bm u - \bm z) \cdot \tilde{\bm u}_\Gamma \, {\rm ds} \right] .
		\end{aligned}		
		\]
Applying integration-by-parts, and using Lemma~\ref{lem:integ}, 
along with $\bm u = 0$ on $\Gamma_{\rm fixed}$ 
and $(\mu \nabla \bm u - pI) \bm n = 0$ on $\Gamma_{\rm out}$, we obtain that 
		\[
		\begin{aligned}	
			L(\bm u, p,\tilde{\bm u}, \tilde{p}, \tilde{\bm u}_\Gamma) 
				&= I(\bm u,p, \bm z) - \left[ \int_\Omega  \mu \nabla \bm u : \nabla \tilde{\bm u}
					-  [ (\bm u \cdot \nabla) \tilde{\bm u} ] \cdot \bm u - p \mbox{div } \tilde{\bm u} + \bm u \cdot \nabla \tilde{p}  \,  {\rm dx} \right. \\
				&\quad + \int_{\Gamma_{\rm in} \cup \Gamma_{\rm fixed}} \tilde{\bm u} \cdot \left(-\mu \nabla \bm u + pI \right) \bm n \, {\rm ds }	
				 -  \int_{\Gamma_{\rm in} \cup \Gamma_{\rm out}} \bm u \cdot \bm n \tilde{p} \, {\rm ds} \\
				&\quad \left.  + \int_{\Gamma_{\rm in}} (\bm u - \bm z) \cdot \tilde{\bm u}_\Gamma \, {\rm ds} 
				+ \int_{\Gamma_{\rm in} \cup \Gamma_{\rm out}} (\bm u \cdot \bm n) (\bm u \cdot \tilde{\bm u}) {\rm ds} \right].
		\end{aligned}		
		\]
Applying integration-by-parts again, we arrive at
		\begin{align}\label{eq:lag}
		\begin{aligned}	
			L(\bm u, p,\tilde{\bm u}, \tilde{p}, \tilde{\bm u}_\Gamma) 
				&= I(\bm u,p,\bm z) - \left[ \int_\Omega \left(- \mbox{div}(\mu \nabla \tilde{\bm u}) + \nabla \tilde{p} \right) \cdot \bm u
					-  [ (\bm u \cdot \nabla) \tilde{\bm u} ] \cdot \bm u - p \mbox{div } \tilde{\bm u}   \,  {\rm dx } \right. \\
				&\quad +  \int_{\Gamma_{\rm in} \cup \Gamma_{\rm fixed}} \tilde{\bm u} \cdot \left(-\mu \nabla \bm u + pI \right) \bm n \, {\rm ds }	
		+  \int_{\Gamma_{\rm in} } \bm u \cdot ( \mu \nabla \tilde{\bm u} - \tilde{p} I)  \bm n \, {\rm ds } \\
		&\quad +  \int_{\Gamma_{\rm out}} \bm u \cdot (\mu \nabla \tilde{\bm u} - \tilde{p} I) \bm n \, {\rm ds }  \\
		&\quad + \left.  \int_{\Gamma_{\rm in}} (\bm u - \bm z) \cdot \tilde{\bm u}_\Gamma \, {\rm ds } 
		+ \int_{\Gamma_{\rm in} \cup \Gamma_{\rm out}} (\bm u \cdot \bm n) (\bm u \cdot \tilde{\bm u}) {\rm ds} \right].
	\end{aligned}		
	\end{align}		

In view of \eqref{eq:Lagopt}, taking a variation of $L$ with respect 
to $(\bm u, p)$  and setting it equal to zero, we obtain the adjoint equation
	\begin{equation}\label{eq:adj}
	\begin{aligned}
		 - \mbox{div}(\mu \nabla \tilde{\bm u}) - (\bm u \cdot \nabla) \tilde{\bm u} - (\nabla \tilde{\bm u})^\top \bm u
			+ \nabla \tilde{p} &= I_{\bm u}(\bm u, p, \bm z) \quad \mbox{in } \Omega \\
		\mbox{div } \tilde{\bm u} &= - I_{p}(\bm u, p, \bm z) \quad \mbox{in } \Omega \\
		\tilde{\bm u} &= \bm 0 \quad \mbox{on } \Gamma_{\rm in} \cup \Gamma_{\rm fixed} \\
		(\mu \nabla \tilde{\bm u} - \tilde{p} I) \bm n &= - \left[ (\bm u \cdot \tilde{\bm u}) \bm n 
			+ (\bm u \cdot \bm n) \tilde{\bm u} \right] \quad \mbox{on } \Gamma_{\rm out}  .
	\end{aligned}	
	\end{equation}
We note the compatibility condition:
	\[
		\tilde{\bm u}_\Gamma = - ( \mu \nabla \tilde{\bm u} - \tilde{p} I )\bm n 
		- (\bm u \cdot \tilde{\bm u}) \bm n - (\bm u \cdot \bm n) \tilde{\bm u} 
		= - ( \mu \nabla \tilde{\bm u} - \tilde{p} I )\bm n 
		\quad \mbox{on } \Gamma_{\rm in} ,
	\]	
where in the last equality we used the fact that 
$\tilde{\bm u} = \bm 0$ on $\Gamma_{\rm in}$. 
We notice that, if $I$ is independent of $p$, then we obtain the 
standard incompressibility condition for $\tilde{\bm u}$ in \eqref{eq:adj}. 
Finally, the required variation of $I$ with respect to $\bm z$ is given by 
	\begin{equation}\label{eq:varJ}
	\begin{aligned}
		D_{\bm z} I(\bm u, p,\bm z) &= I_{\bm z}(\bm u, p,\bm z) - \left[ (\mu\nabla \tilde{\bm u} - \tilde{p} I) \bm n 
		+ (\bm u \cdot \tilde{\bm u}) \bm n + (\bm u \cdot \bm n) \tilde{\bm u} 	  \right]  \quad \mbox{on } \Gamma_{\rm in} \\
		&= I_{\bm z}(\bm u, p,\bm z) - \left[ (\mu\nabla \tilde{\bm u} - \tilde{p} I) \bm n  \right]  \quad \mbox{on } \Gamma_{\rm in} \, ,
	\end{aligned}	
	\end{equation}
where we have again used the fact that $\tilde{\bm u} = \bm 0$ on $\Gamma_{\rm in}$. Note that if the clinical measure $I$ is
not a function of the control variable (in this case the inflow velocity), for a channel with constant flow in the normal direction $\bm n$ (i.e. $\mu \nabla \tilde{\bm u} \cdot \bm n = 0$) the sensitivity reverts to (recall that $\mathcal{I}$ is the reduced objective)
\begin{equation}\label{eq:varin}
		D_{\bm z} \mathcal{I}(\bm z) = \tilde{p} \bm n
         \quad \mbox{on } \Gamma_{\rm in} \, .
\end{equation}
i.e. \emph{the sensitivity to inflow velocities is the adjoint pressure}.

\subsubsection{Sensitivity to Changes in Inflow Position}

Consider next the variation of the Lagrangian $L$ given in \eqref{eq:lag} with
respect to the normal $\bm n$. We recall that after simplifications, we have
\[
     L(\bm u, p,\tilde{\bm u}, \tilde{p}, \tilde{\bm u}_\Gamma) 
     = I(\bm u,p,\bm z)
     - \int_{\Gamma_{\rm in}} (\bm u - \bm z)
       \left[( \mu \nabla \tilde{\bm u} - \tilde{p} I )\bm n \right] \, .
\]
Then 
\[
\begin{aligned}
    D_{\bm n} L(\bm u, p,\tilde{\bm u}, \tilde{p}, \tilde{\bm u}_\Gamma)  \bm h 
    &= D_{\bm n} I (\bm u,p,\bm z) \bm h 
     - \int_{\Gamma_{\rm in}} D_{\bm n} \left[ (\bm u - \bm z)
       \left( ( \mu \nabla \tilde{\bm u} - \tilde{p} I )\bm n \right) \right] \bm h \\
    &= D_{\bm n} I (\bm u,p,\bm z) \bm h \\
    &\quad - \int_{\Gamma_{\rm in}} (D_{\bm n} \bm u \bm h ) \left[
       \left( ( \mu \nabla \tilde{\bm u} - \tilde{p} I )\bm n \right) \right] 
       + (\bm u - \bm z) D_{\bm n} \left[ 
       \left( ( \mu \nabla \tilde{\bm u} - \tilde{p} I )\bm n \right) \right] \bm h \\
    &= D_{\bm n} I(\bm u,p,\bm z) \bm h 
     - \int_{\Gamma_{\rm in}} (D_{\bm n} \bm u \bm h ) \left[
       \left( ( \mu \nabla \tilde{\bm u} - \tilde{p} I )\bm n \right) \right] \, ,
\end{aligned}    
\]
where, in the last step, we have used the fact that $\bm u = \bm z$
on $\Gamma_{\rm in}$. In case, $I$ is independent of $\bm n$, 
we then obtain that 
\[
	D_{\bm n} L(\bm u, p,\tilde{\bm u}, \tilde{p}, \tilde{\bm u}_\Gamma)  \bm h 
	 = - \int_{\Gamma_{\rm in}} (D_{\bm n} \bm u \bm h ) \left[
      	      \left( ( \mu \nabla \tilde{\bm u} - \tilde{p} I )\bm n \right) \right] \, .
\]

\noi
Note that if $\mu \nabla \tilde{\bm u} \cdot \bm n = 0$ (as is often the case) the sensitivity reverts to (recall that $\mathcal{I}$ is the reduced objective)
\begin{equation}\label{eq:varin}
		D_{\bm n} \mathcal{I}(\bm n) = u^n_n \tilde{p} 
  \quad \mbox{on } \Gamma_{\rm in}
\end{equation}

\noi
i.e. \emph{the sensitivity to changes in inflow position is the adjoint pressure multiplied by the normal derivative of the inflow velocity}. 

\subsection{In- and Outflow Boundary Conditions for the Adjoint}

\begin{figure}[h!]
\centering
\includegraphics[width=10.0cm]{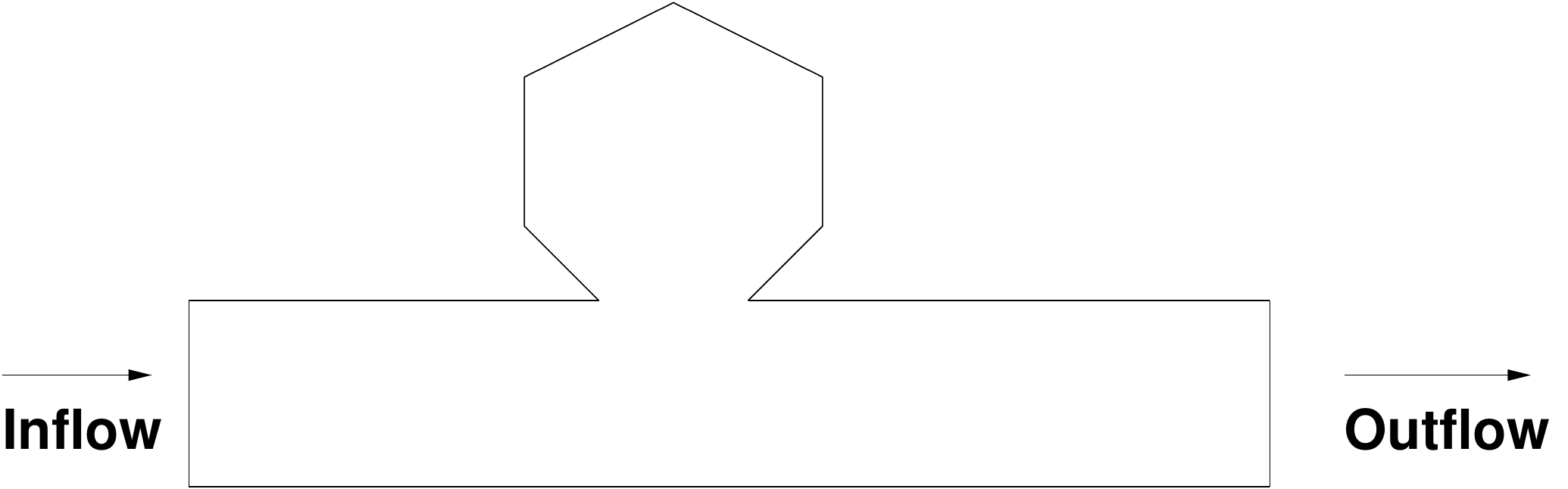}
\caption{\label{f:any}Schematic of Aneurysm}
\end{figure}

Consider the aneurysm shown in Figure~\ref{f:any}.

\ms \noi
For the usual (forward) incompressible Navier-Stokes calculation,
one would prescribe a velocity profile ($\bm u =\bm z$) at
the inflow boundary and the `do nothing' ($(\eta \nabla \bm u - pI) \bm n = 0$)
or pressure boundary condition ($p=p_{ou}$) at the outflow
boundary. This implies letting the pressure `free' at the inflow
and the velocity `free' at the outflow. At the walls the velocity
is zero, i.e. $\bm u |_{\Gamma_{\rm fixed}}=\bm 0$.
Consider now the adjoint problem. The boundary conditions in this 
case are described in \eqref{eq:adj}, i.e., we obtain zero
velocity at the inflow and `do nothing' or prescribed zero adjoint
pressure at the outflow. The adjoint velocity is also zero on the walls.

\section{Numerical Implementation}
\label{s:numimp}

In a strict mathematical sense, the adjoint solver obtained by
discretizing the adjoint partial differential equation should be
as close as possible to the discrete adjoint obtained from
transposing and manipulating the discretization of the forward
problem. In this way `optimize-then-discretize' and
'discretize-then-optimize' are as close as possible.
This was not adopted in the present case. Instead, while
the forward problem was solved for the incompressible Navier-Stokes
equations, the adjoint equations were derived for the
quasi-incompressible Navier-Stokes equations, which for steady flows give the same results. 
Furthermore, while 
the forward problem was integrated to steady state using a fractional
step solver with implicit solution of the viscous terms and the
pressure increments, and edge-based upwinding for the velocities
and 4th order pressure stabilization \cite{RLoehner_2008a},  
the adjoint was discretized
in space using the following scheme, which for each point $i$ in
the mesh is given by: 
$$ \left[\Amat^k \right]^T_i M_i \Grd^k(\tilde{\bm u})_i
 + \Bmat^T (\mu_i + \mu_j) K_{ij} ({\tilde{\bm u}}_i-{\tilde{\bm u}}_j) +
   M_i I^{\Omega}_{\bm u} + D_i = 0 \, , \eqno(**) $$
where $\Amat, M_i, \Grd^k, K_{ij}, D_i$ denote the Jacobians of the advective fluxes, lumped mass-matrix,
discrete gradient in direction $k$, Laplacian edge-based coefficients
and damping vector, and
$$ \Grd^k(\tilde{\bm u})_i = C^k_{ij} ( {\tilde{\bm u}}_i + {\tilde{\bm u}}_j ) 
\, , $$
where $C^k_{ij}$ are the edge-based coefficients for the gradient (see \cite{RLoehner_2008a}, Chapter~20).
Furthermore
$$ D_i = - \lambda^{(ij)} \left[ \tilde{\bm u}_i - \tilde{\bm u}_j
+ {\beta \over 2} \lvec_{ij} \cdot 
  ( \Grd(\tilde{\bm u})_i + \Grd(\tilde{\bm u})_j ) \right]
\, , $$
where $c$ is the speed of sound, $\tilde{p}$ the adjoint
pressure, $\lambda=|\bm u|+c$ the 
maximum eigenvalue of the system and
$0< \beta < 1$ denotes a pressure sensor function of the form
\cite{peraire19923d}.
$$
\beta = 1 -
{ {\tilde{p}_i-\tilde{p}_j  +  0.5 \lvec_{ij} \cdot (\Grd(\tilde{p})_i + \Grd(\tilde{p})_j)|} \over
  {|\tilde{p}_i-\tilde{p}_j| + |0.5 \lvec_{ij} \cdot (\Grd(\tilde{p})_i + \Grd(\tilde{p})_j)|} }
~~. \eqno(9.4) $$

\noi
For $\beta=0,1$, second and fourth order damping operators
are obtained respectively.
Several other forms are possible for the sensor function
$\beta$ \cite{mestreau1993tgv}.

Although this discretization of the adjoint Euler fluxes looks like a blend of second and fourth order dissipation, it has no adjustable parameters. 
Defining $\bm U = (\bm u, p)$, $\tilde{\bm U} = (\tilde{\bm u}, \tilde{p})$ 
Eqn.(**) may be re-written as
$$ \Rvec(\bm U, \tilde{\bm U})=0 \, , $$
the system re-written as an unsteady equation of the form:
$$ \tilde{\bm U}_{,\tau} + \Rvec(\bm U,\tilde{\bm U}) = 0 \, , $$
and integrated in pseudo-time $\tau$ via a classic explicit multistep
Runge-Kutta \cite{jameson1981numerical}.

\section{Numerical Examples}

We will focus on two main examples. At first, we consider Poisuille flow
through a channel in Section~\ref{s:Poiseuille}. Remarkably enough, we 
are able to derive the explicit expressions for all the quantities, such
as solution to the state equation, adjoint equation and sensitivities, see 
Appendix~\ref{app:analytical}. These theoretical results are also confirmed 
by numerical results. In Section~\ref{s:aneu}, we focus on a realistic
aneurysm scenario, where we truly see the benefits of the proposed sensitivity
approach. 

\subsection{Poiseuille Flow}
\label{s:Poiseuille}

The 2-D channel flow provides a good test to verify the implementation
of the forward and adjoint solvers. The domain considered is of
dimension $0.0 \le x \le 0.5$, $-0.05 \le y \le 0.05$ and
$-0.005 \le z \le 0.005$. A parabolic inflow with maximum velocity
of $u_{max}=1.0$ was prescribed. 
The velocity at the top and bottom 
walls ($y_{min}, y_{max}$) was prescribed to zero, and the velocity
in the $z$-direction was prescribed to zero for the back and front
walls ($z_{min}, z_{max}$). The other relevant parameter 
is $\mu=0.01$. Two `clinically relevant measures'
(i.e. cost functions) were considered: kinetic energy
$I = \frac12 \int_{\Omega} \rho {\bm u}^2 \mbox{ dx}$ 
and vortical energy 
$I = \frac12 \int_{\Omega} \rho |\nabla \times \bm u|^2 \mbox{ dx}$. 
We set $\rho = 1.0$ in our experiments. 
The derivation of the exact solutions for the adjoint 
equations for these cost functions may be found in Appendix~\ref{app:analytical}.
Let $\bm u = (u,v,w)^\top$, then the $x$-component of $\bm u$ is given by:
$$ u = \left[ 1 - {4 \over H^2} y^2 \right] u_0 \, , $$
where $u_0=u_{max}$ and $H$ is the total height of
the channel, i.e. $y_{max}=-y_{min}=H/2$. We thus
obtain 
\[
\partial_y u = - {{8 u_0} \over H^2} y, \quad 
\partial_{yy} u = - {{8 u_0} \over H^2}, \quad 
\partial_x p = - {{8 \mu u_0} \over H^2} .
\]

\begin{figure}[h!]
    \centering
    \includegraphics[width=10.0cm]{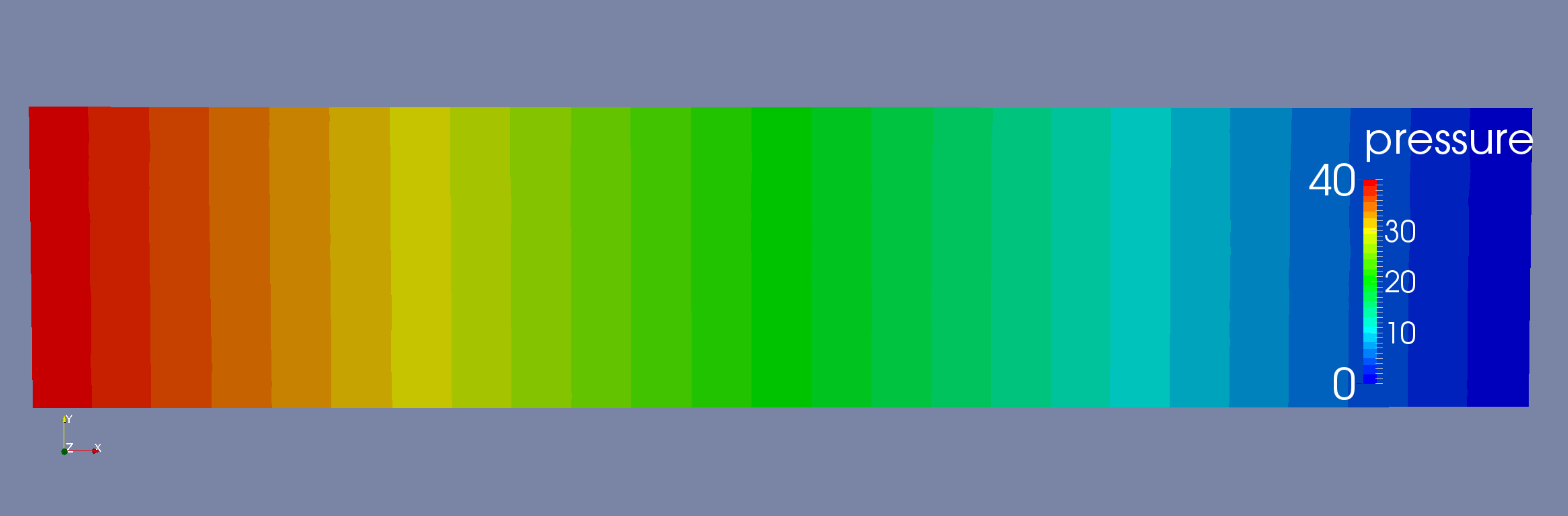}
    \caption{Poiseuille Flow: Pressure}
    \label{f:ppres}
\end{figure}

\begin{figure}[h!]
    \centering
    \includegraphics[width=10.0cm]{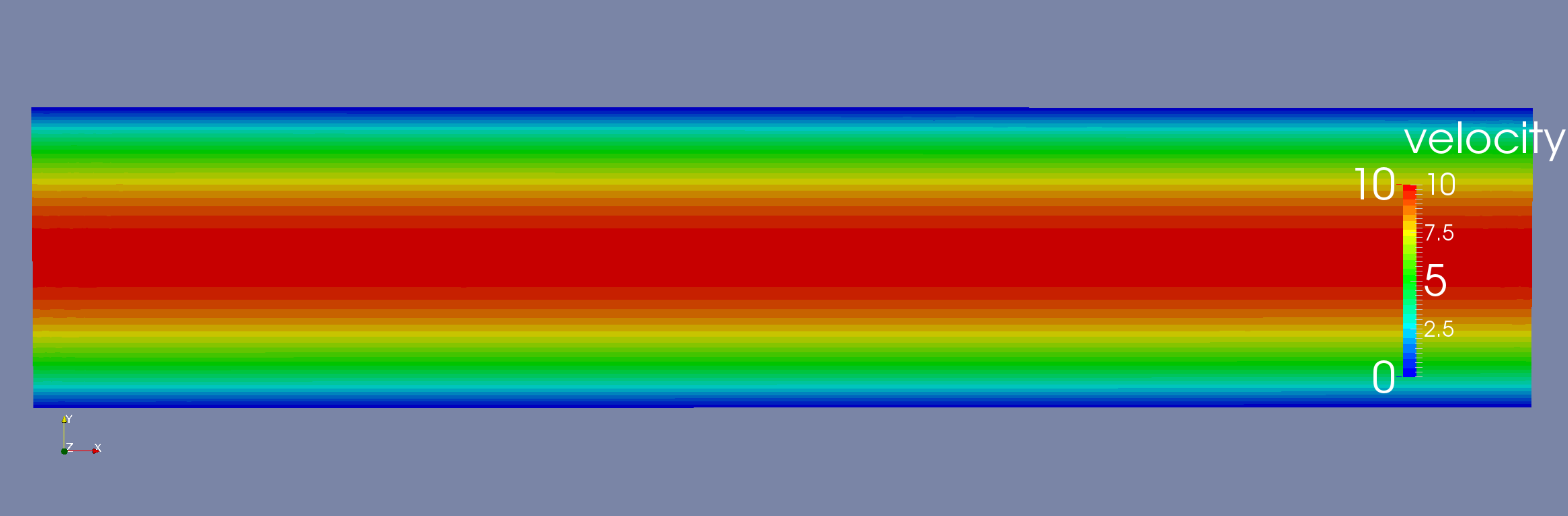}
    \caption{Poiseuille Flow: Velocity Magnitude}
    \label{f:pvelo}
\end{figure}

\begin{figure}[h!]
    \centering
    \includegraphics[width=10.0cm]{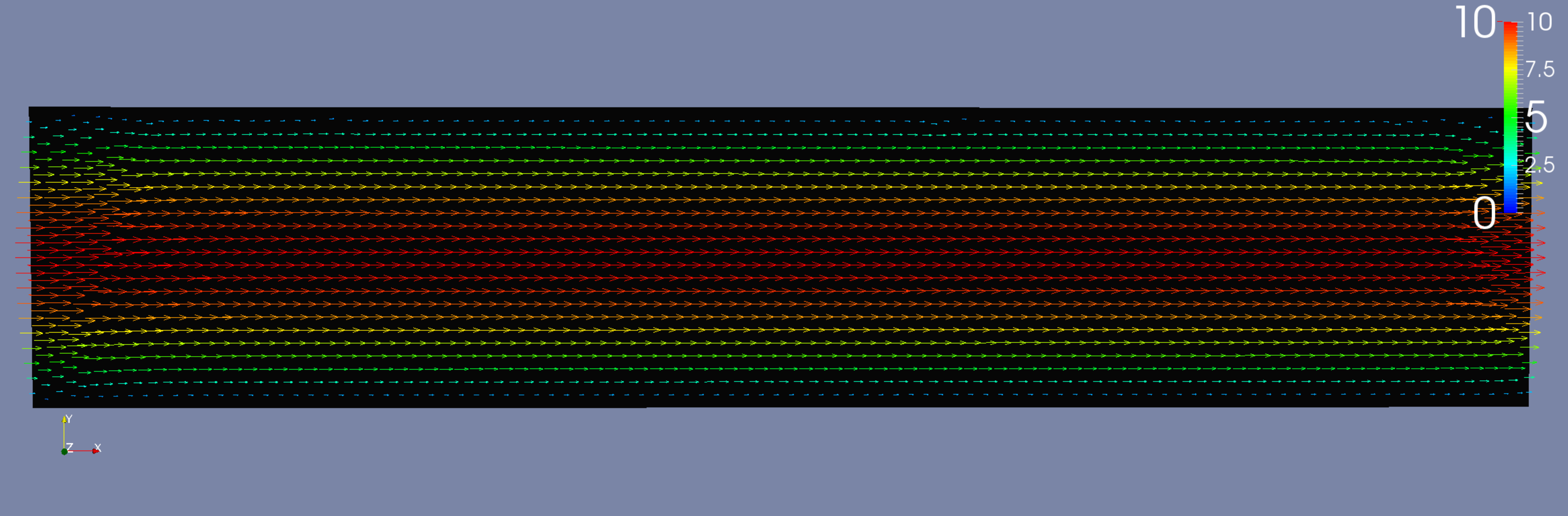}
    \caption{Poiseuille Flow: Velocity}
    \label{f:pveloarrow}
\end{figure}

The pressure, velocity magnitude, and velocity vectors
are shown in 
Figures~\ref{f:ppres}-\ref{f:pveloarrow}. 

\subsubsection{Kinetic Energy}

\begin{figure}[h!]
    \centering
    \includegraphics[width=10.0cm]{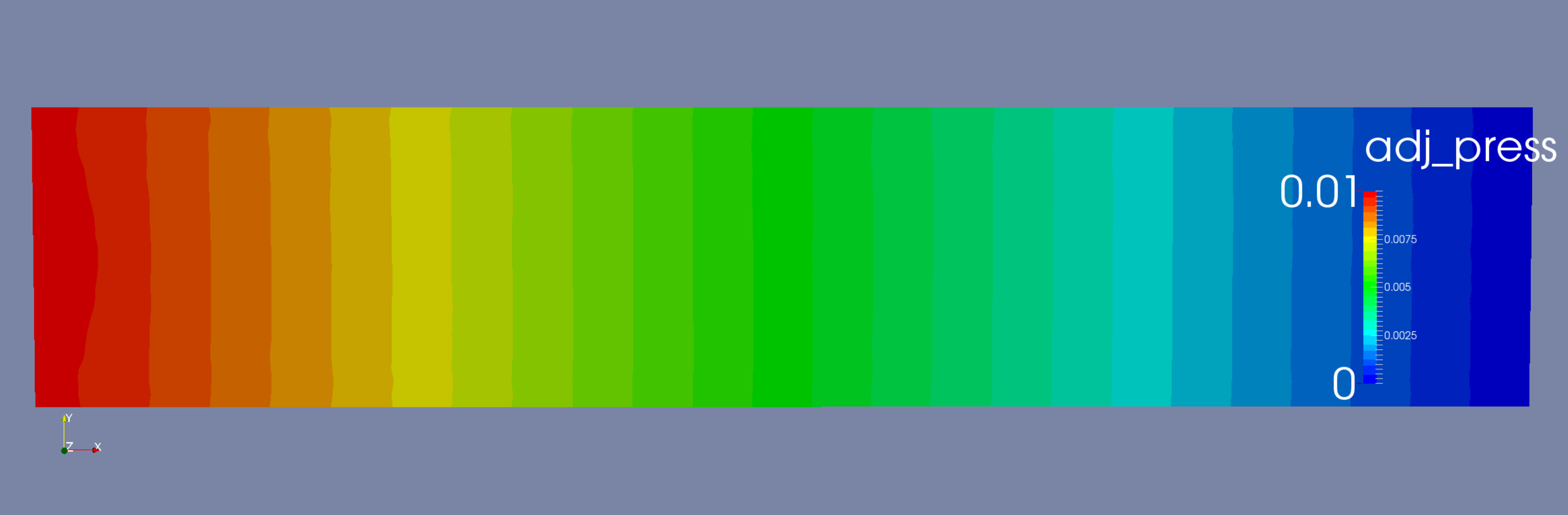}
    \caption{Poiseuille Flow: Adjoint Pressure}
    \label{f:padjpres}
\end{figure}

\begin{figure}[h!]
    \centering
    \includegraphics[width=10.0cm]{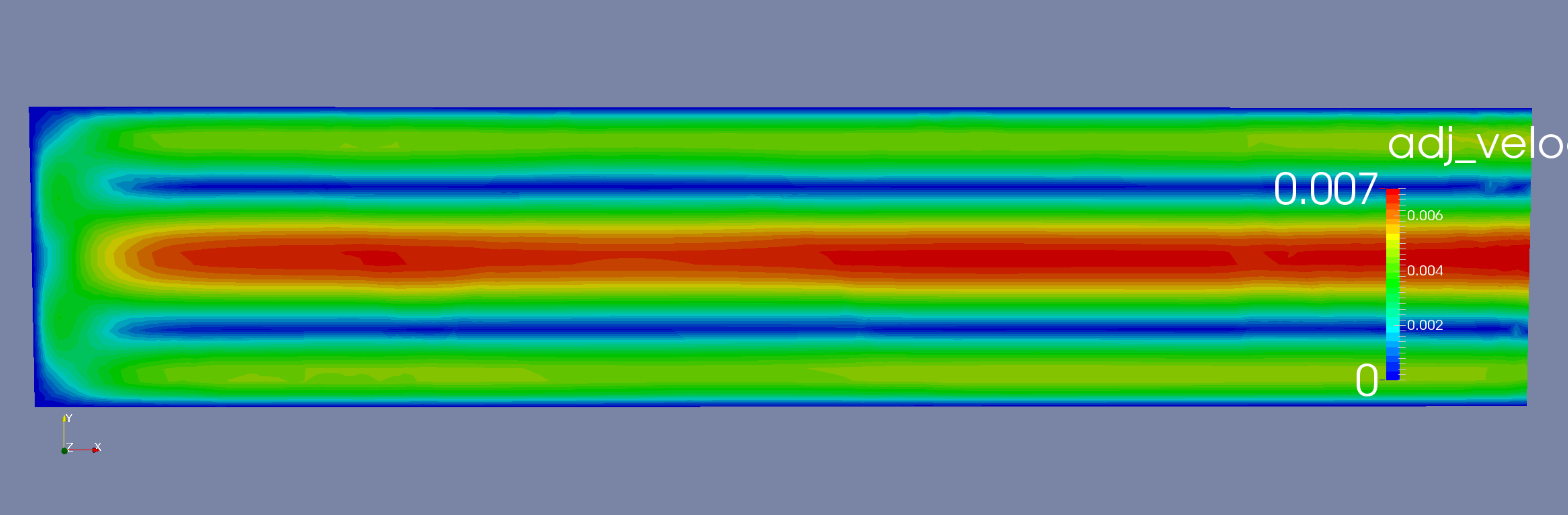}
    \caption{Poiseuille Flow: Magnitude of Adjoint Velocity. Here the cost function 
        is Kinetic Energy.}
    \label{f:pveloadj}
\end{figure}

\begin{figure}[h!]
    \centering
    \includegraphics[width=10.0cm]{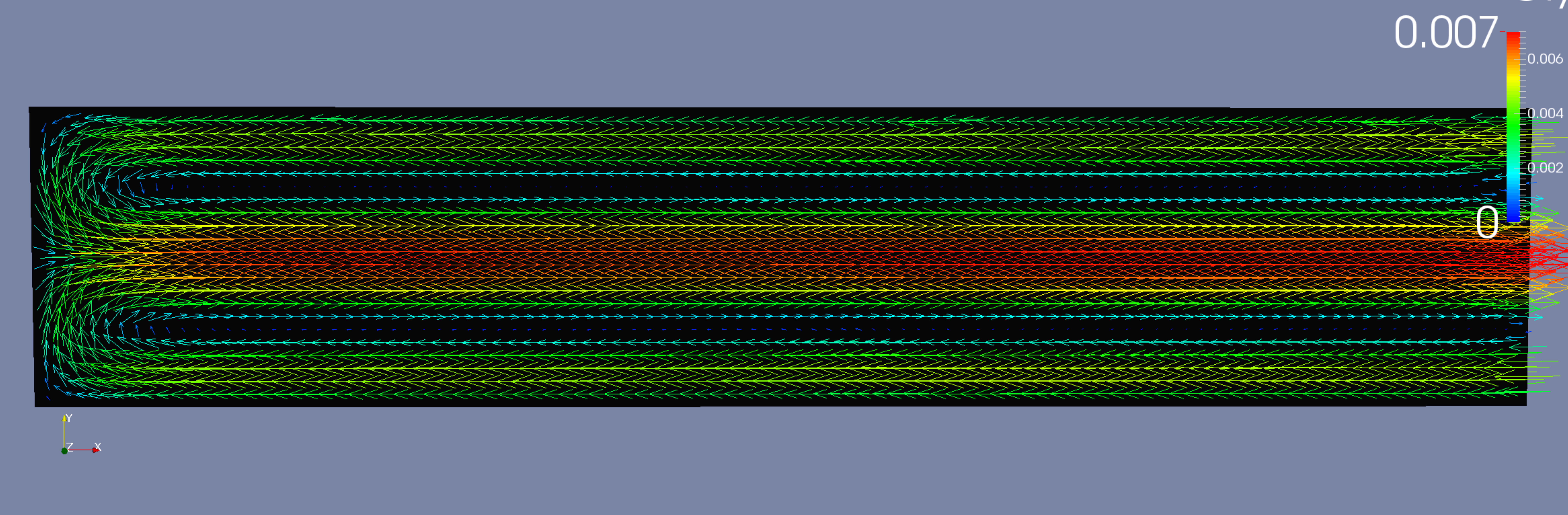}
    \caption{Poiseuille Flow: Adjoint Velocity. Here the cost function 
        is Kinetic Energy.}
    \label{f:pveloadjarrow}
\end{figure}

Consider the cost function 
$$ I = \frac12 \int \rho |\bm u|^2 \mbox{ dx} \, , $$
implying
$$ I_{u} = \rho u . $$
As can be seen in Appendix~1, the adjoint pressure for this
cost function is:
$$ \partial_x \ptilde =  {4 \over 5} \rho u_0 ~~, $$
i.e. the gradient of the adjoint pressure is also constant and
linearly dependent of $u_0$. The results obtained are shown
in Figures~\ref{f:padjpres}-\ref{f:pveloadjarrow}.

\subsubsection{Vortical Energy}

The cost function is given by
$$ I = { 1 \over 2} \int \rho \left| \nabla \times \bm u \right|^2 d\Omega
                                       ~~. $$
For the 2-D channel ($u=u(y),v=0,w=z$)

$$ ( \nabla \times \bm u )^2 = \left( \partial_y u \right)^2 ~~, $$

\noi
so that
$$ I_{,u} = \rho u_{,y} (u_{,y})_{,u} = - \rho u_{,yy}
          = - {\rho \over \mu} p_{,x} ~~
          = {{8 \rho u_0} \over H^2} ~~,
$$
i.e. constant. As can be seen in Appendix~1, the adjoint
velocities and pressure are given by:
$$ \utilde(x,y)=0 ~~, ~~ \vtilde(x,y)=0 ~~,
   -\ptilde = {\rho \over \mu} p ~~. $$

\begin{figure}[h!]
    \centering
    \includegraphics[width=10.0cm]{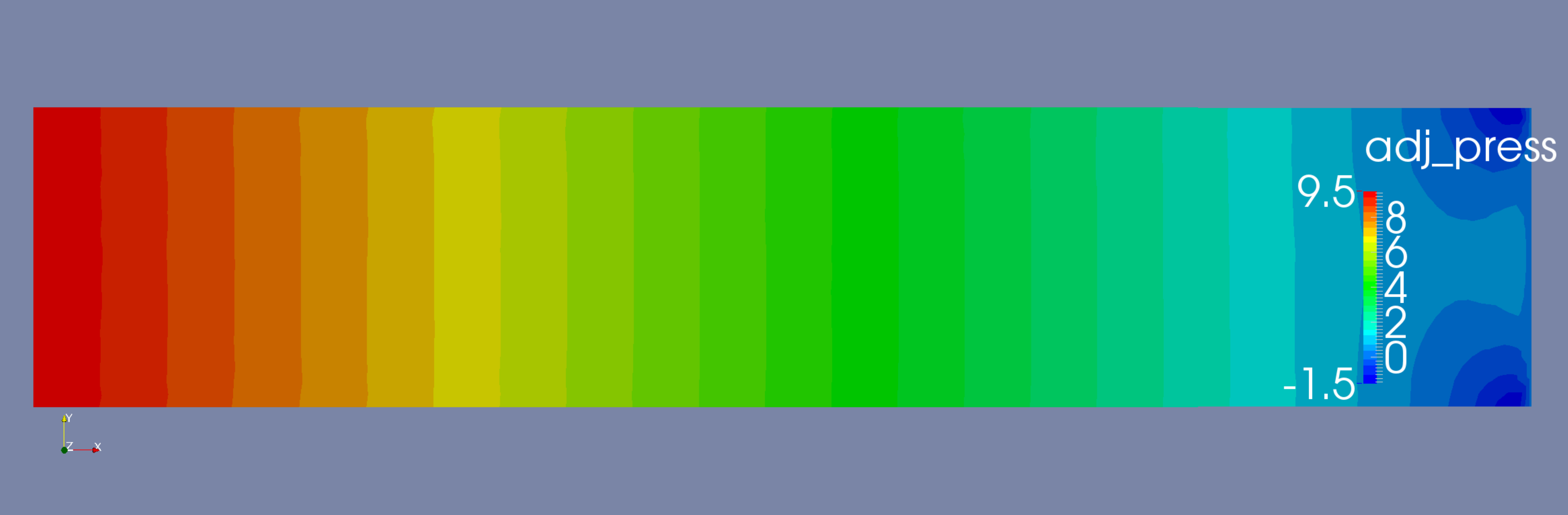}
    \caption{Poiseuille Flow: Adjoint Pressure. Cost Function: Vortical Energy.}
    \label{fig:my_label}
\end{figure}

\subsection{Aneurysm with Simple Flow Pattern}
\label{s:aneu}

As an example, we include an aneurysm with simple flow pattern.
The geometry and discretization may be discerned from
Figures~11a-c which show the surface triangulation,
pressure and magnitude of the velocity.
The region for the source-terms of the adjoint is shown in Figure 12~a and the adjoint pressure, as well as the magnitude of the adjoint velocities obtained in Figures~12~b,c.
The adjoint velocites can also be seen in Figures~13~a,b. Note the effect of the source-term that pushes the adjoint flow and forms a double vortex.

\begin{figure}[h!]
    \centering
    \includegraphics[width=5.0cm]{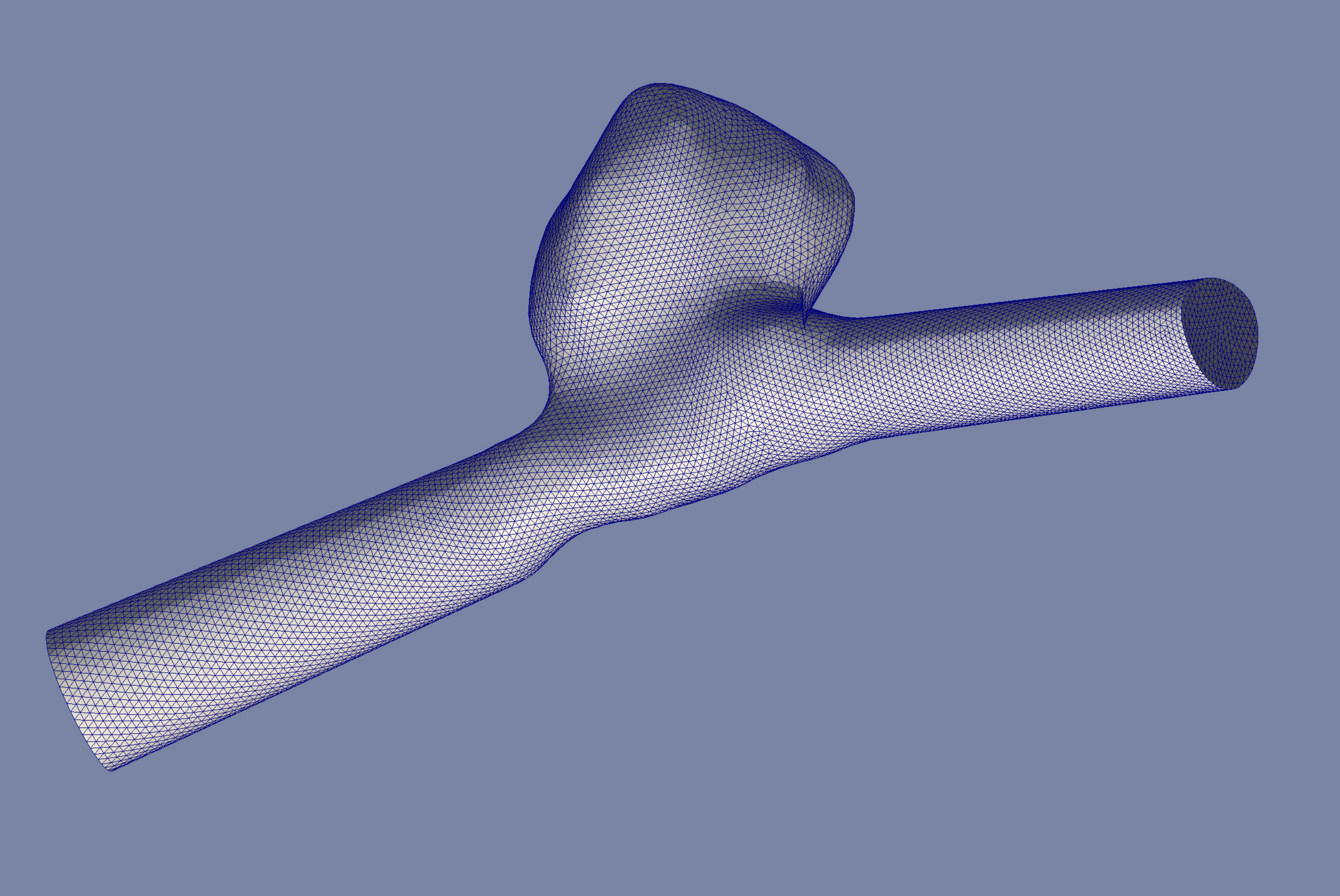}
    \includegraphics[width=5.0cm]{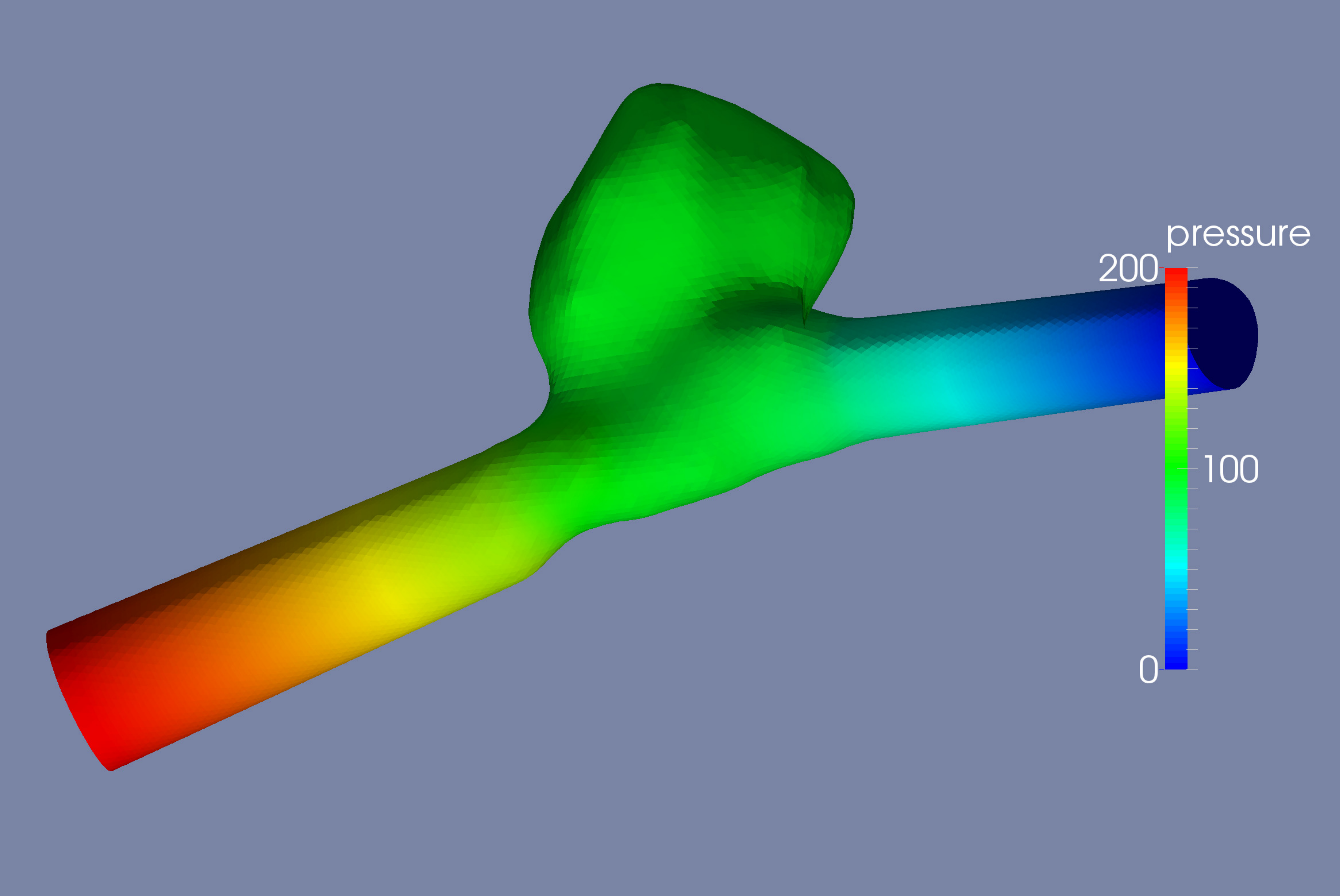}
    \includegraphics[width=5.0cm]{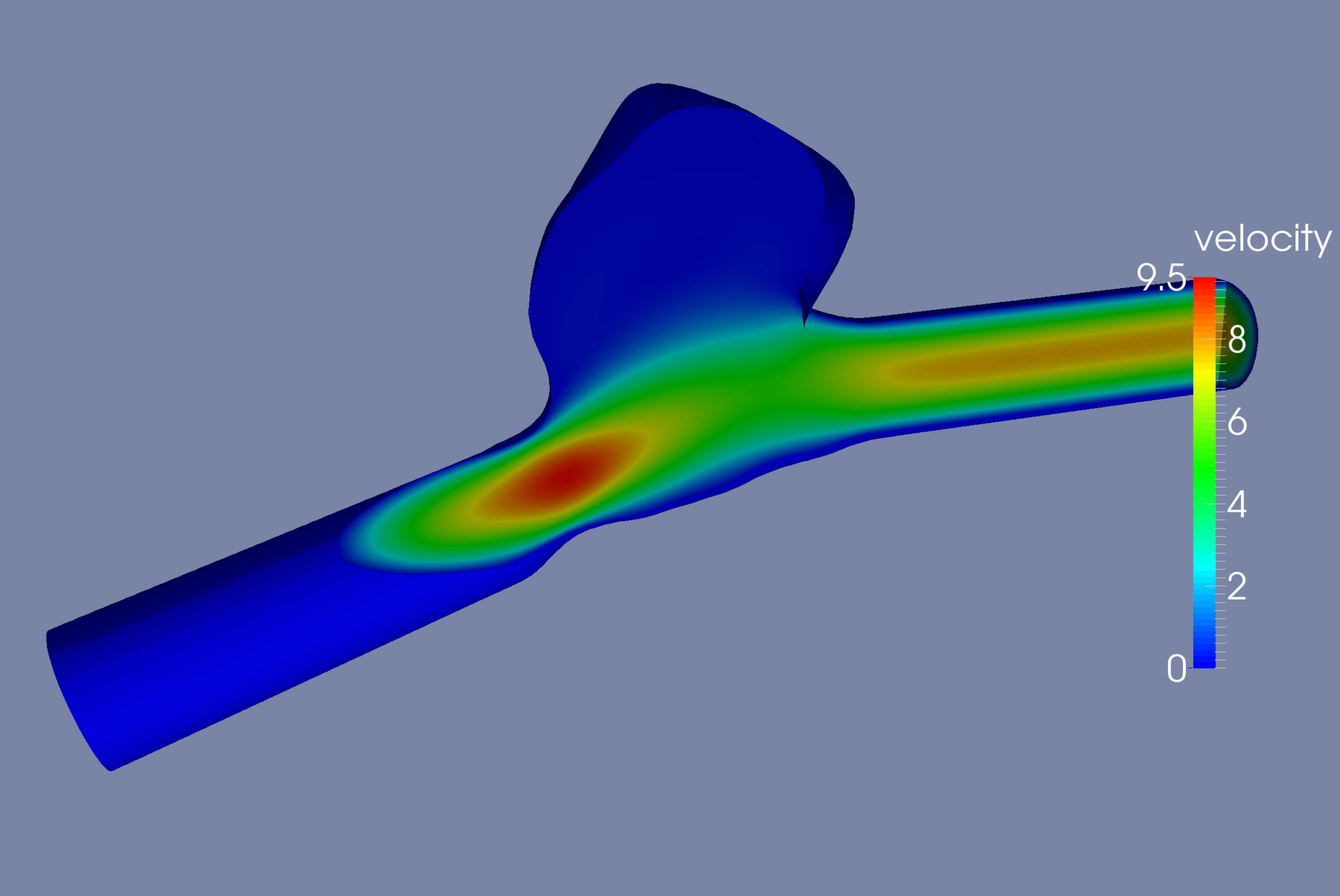}
    \caption{a,b,c~~Aneurysm: Surface Triangulation, Surface Pressure and Magnitude of Velocity in Cut Plane}
    \label{fig:my_label}
\end{figure}

\begin{figure}[h!]
    \centering
    \includegraphics[width=5cm]{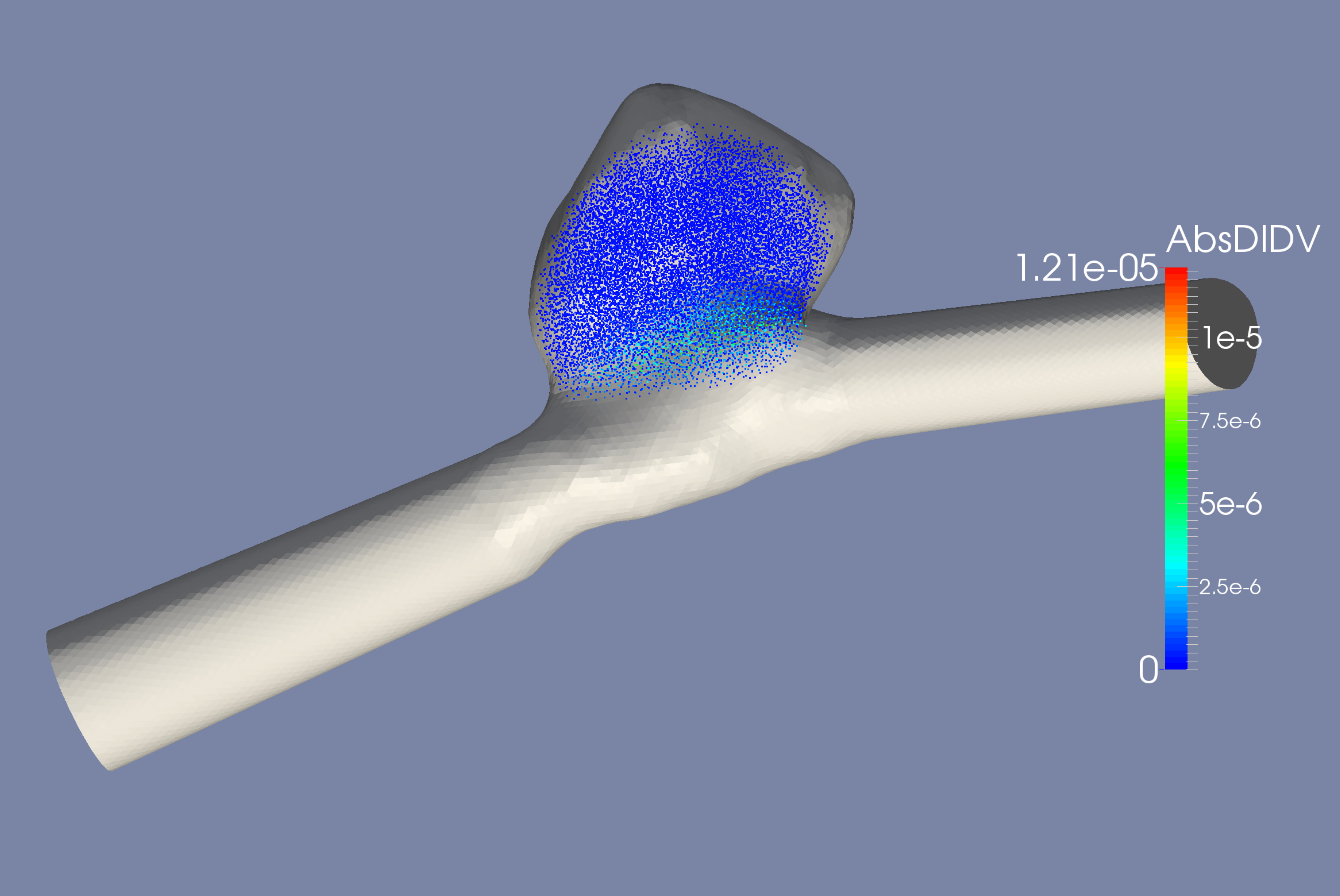}
    \includegraphics[width=5cm]{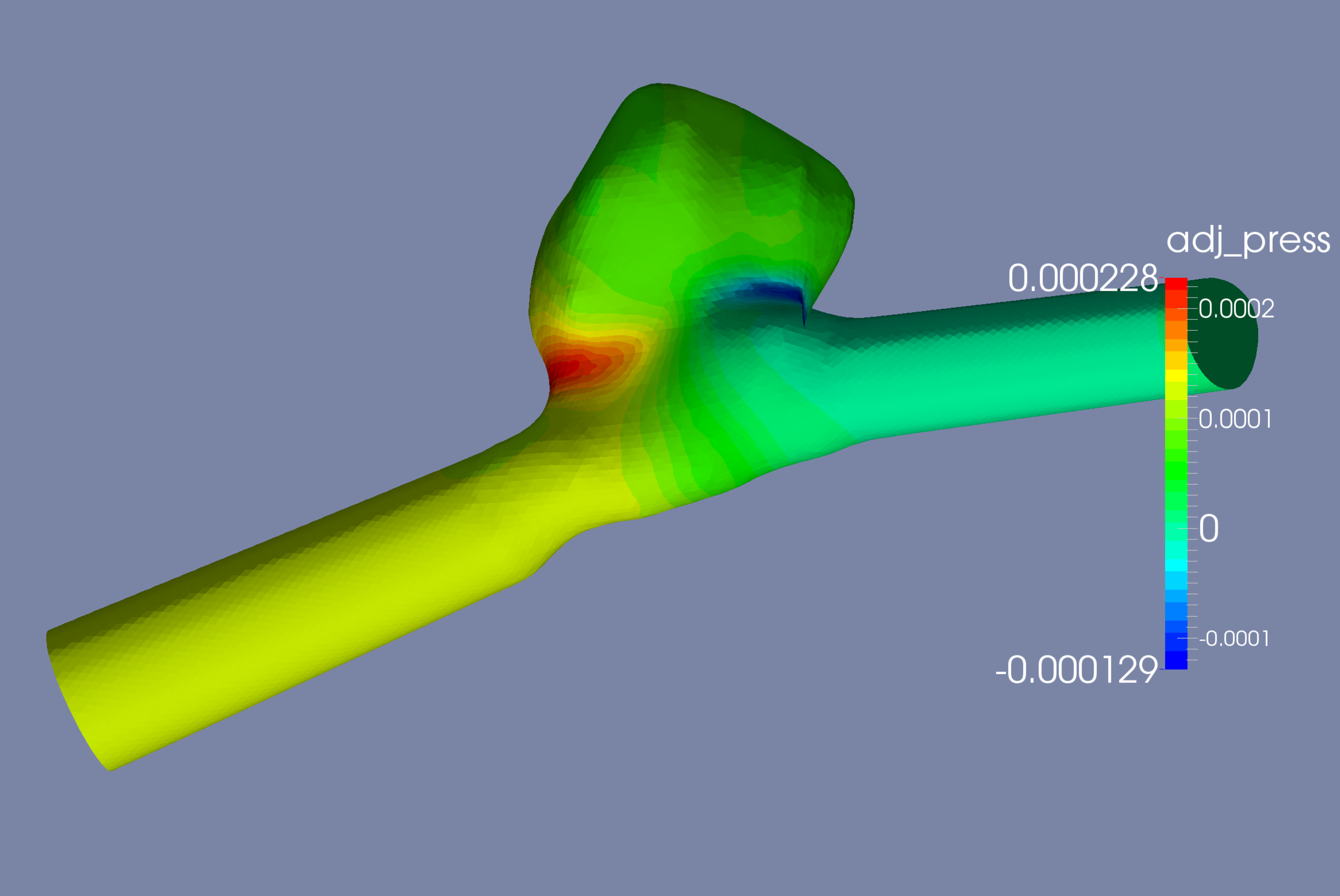}
    \includegraphics[width=5cm]{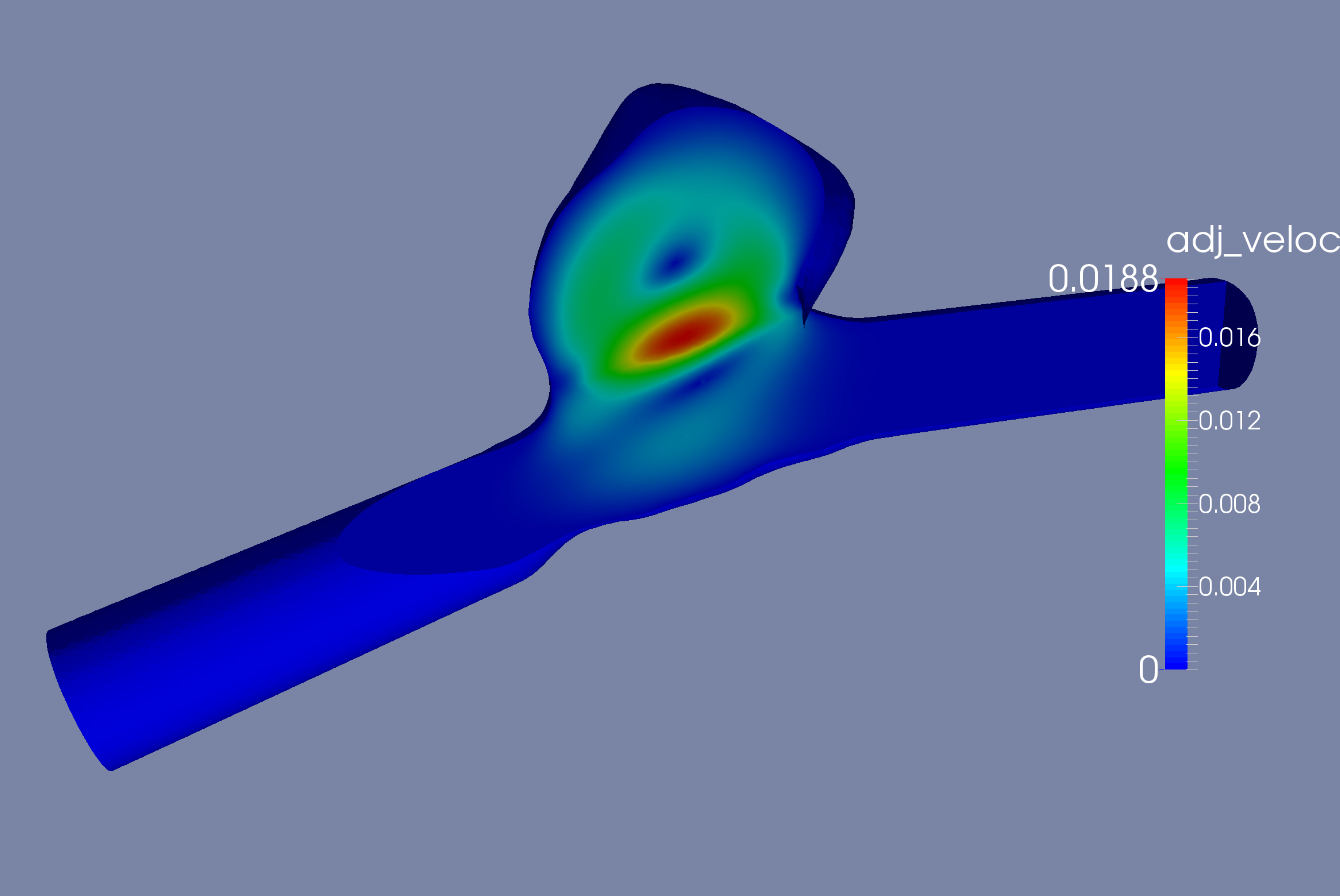}
    \caption{a,b,c~~Aneurysm: Source, Adjoint Pressure and Magnitude of Adjoint Velocity in Cut Plane}
    \label{fig:my_label}
\end{figure}

\begin{figure}[h!]
    \centering
    \includegraphics[width=5cm]{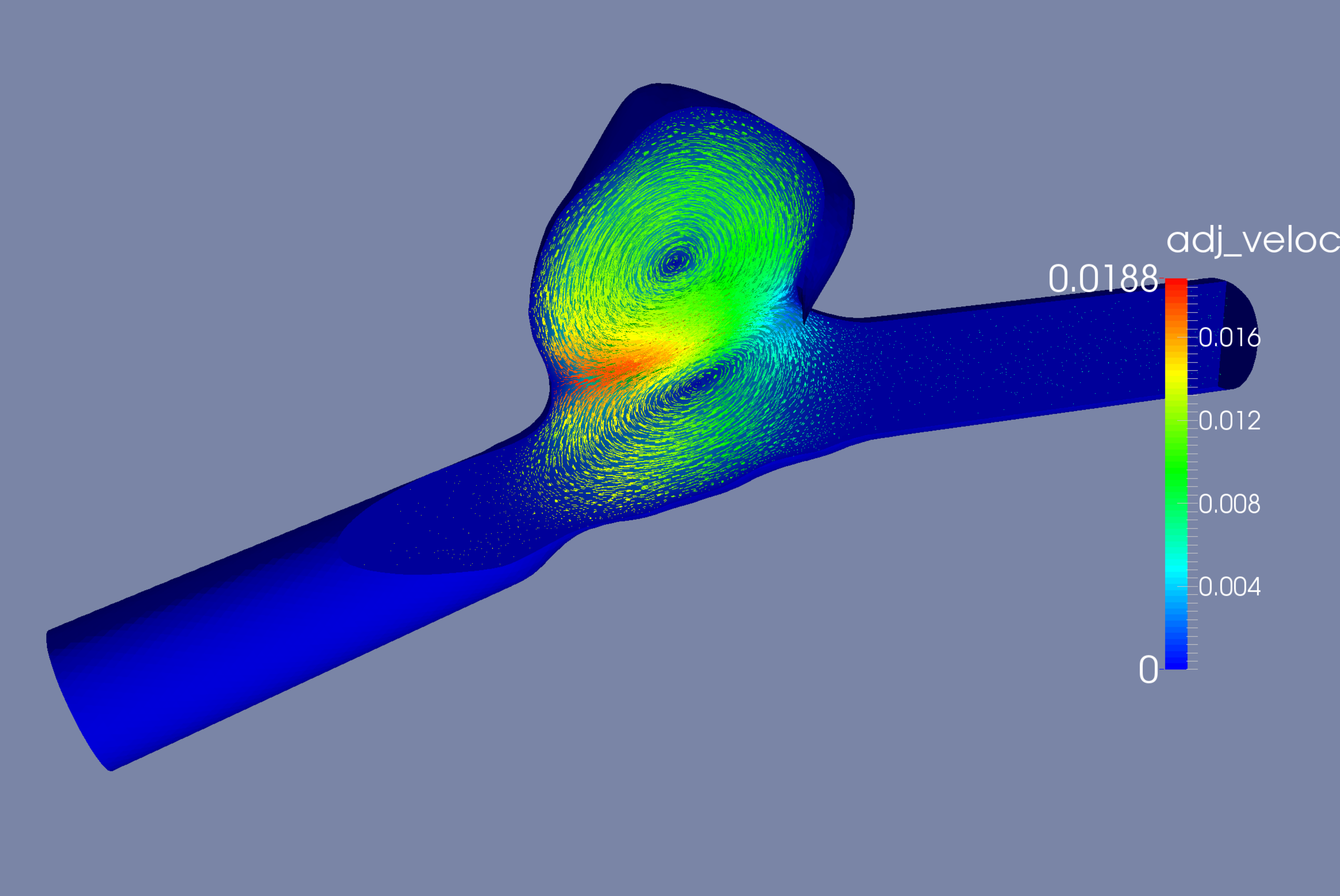}
    \includegraphics[width=5cm]{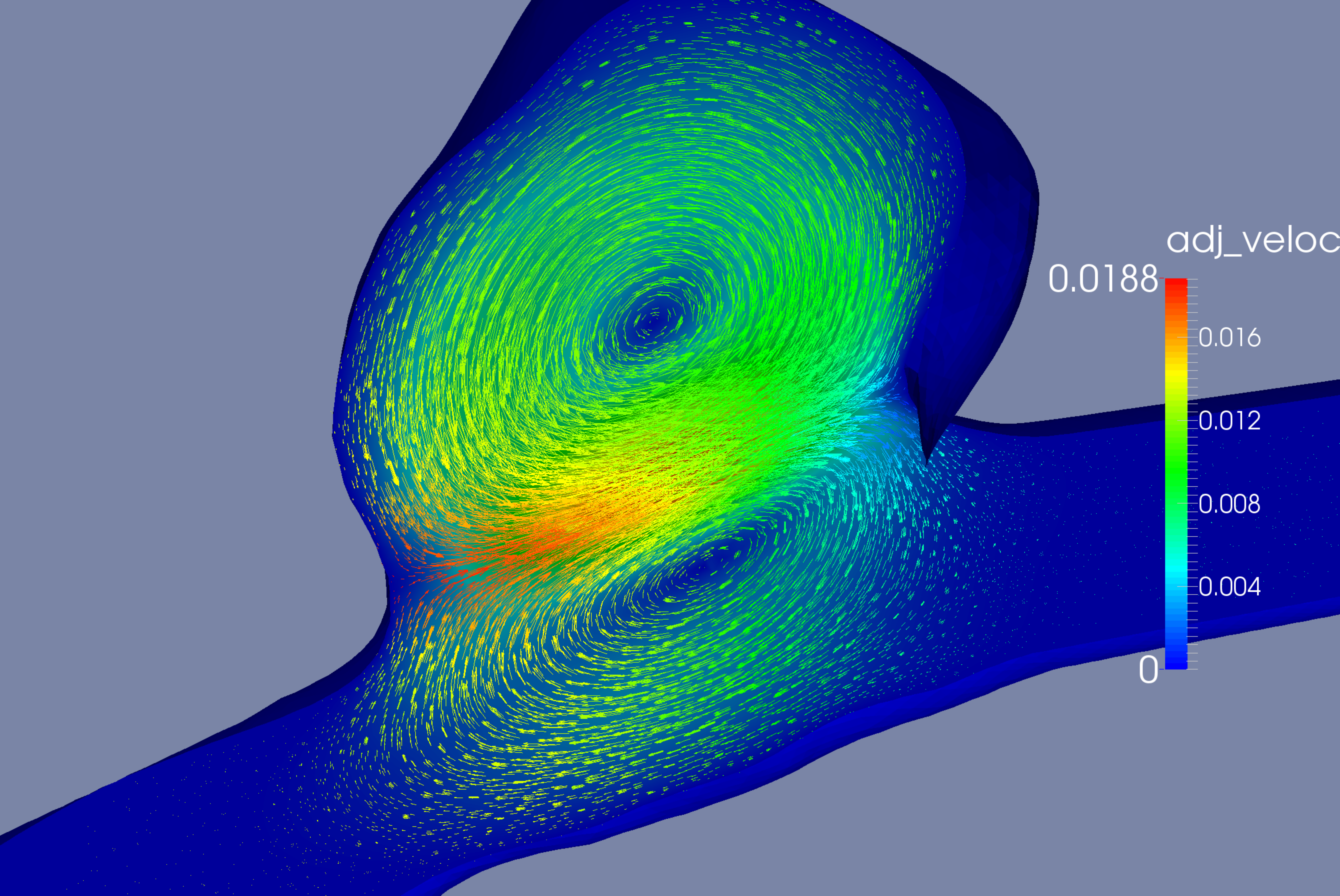}
    \caption{a,b~~Aneurysm: Adjoint Velocity in Cut Plane}
    \label{fig:my_label}
\end{figure}

\section{Conclusions and Outlook}
\ms \noi
The use of adjoint solvers to assess the sensitivity of incomplete boundary (inflow, geometry) information has been considered. The results of this investigation indicate that the sensitivity of clinical measures or other flow features that are inside the flow domain with respect to inflow velocity is proportional to the adjoint pressure, while the sensitivity with respect to inflow geometry is given by the product of the adjoint pressure and the normal derivative of the inflow velocity. Thus, the adjoint pressure may be a good indicator to see if the inflow boundary of haemodynamic cases is far enough from the region of interest so that errors can be avoided.
\noi
The use of adjoint solvers is not unproblematic. Unlike running a series of cases, varying inflow profiles and geometry, and seeing their influence on many clinically relevant measures, adjoints require a different run for
each of the clinical measures.

\appendix
\section{Appendix 1: Analytical Expressions for Poiseuille Flow}
\label{app:analytical}

\subsection{Exact Forward Solution}

Let us consider a long 2-D channel of length $0 \le x \le L$ and
width $-H/2 \le y \le H/2$ with incompressible viscous flow.
Let $\bm u = (u, v , w)^\top$, then 
the equation for the $x$-velocity $u$ is given by:
$$ u \partial_x u + v \partial_y u + \partial_x p = 
   \mu \Delta u  \, . $$
Assuming a constant velocity profile in $x$, i.e. $u=u(y)$ and laminar
flow with $v=0$,  the solution is the Poiseuille solution, given by:
\begin{equation}\label{eq:u}
	u = \left[ 1 - {4 \over H^2} y^2 \right] u_0 \, , 
\end{equation}	
where $u_0$ is the maximum velocity at the center of the channel, and the channel extends in height from $-H/2 \le y \le H/2$, implying
$$ \partial_y u = - {{8 u_0} \over H^2} y \, , $$
and
$$ \partial_{yy} u = - {{8 u_0} \over H^2} \, , $$
so that the constant pressure gradient is given by:
$$ \partial_x p = - {{8 \mu u_0} \over H^2} \, , $$
where we have used the fact that $\partial_x u = \partial_{xx} u = 0$. 
The average velocity is then:
$$ {\overline u} = { 1 \over H} \int_{-H/2}^{H/2} u \mbox{ dy} = {2 \over 3} u_0 \, . $$
%

\subsection{Adjoint Equations}

The equation for the adjoint $x$-velocity $\utilde$ is given by:

$$ - u \partial_{x} \utilde - v \partial_y \utilde + \partial_x \ptilde = 
   \mu \Delta \utilde_{,xx}  + I_{u} $$

\noi
Here $I$ is the cost function. For the channel $u$ is given by
\eqref{eq:u} and $v=0$.

\bs \noi
\ub{Kinetic Energy}: 
\ms \noi
If the cost function is given by the kinetic energy
$$ I = { 1 \over 2} \int \rho |\bm u|^2 \mbox{ dx} \, , $$
then
$$ I_{u} = \rho u \, .$$ 
Assuming a long channel with no change in $x$ of the variables, the
equation for the adjoint $x$-velocity $\utilde$ simplifies to:
$$ \partial_x \ptilde = \mu \partial_{yy} \utilde 
  + \rho u_0 \left[ 1 - {4 \over H^2} y^2 \right] \, . $$
Assuming furthermore that $\partial_x \ptilde$ is constant, and applying the
boundary conditions $\utilde=0$ for $y=-H/2$ and $y=H/2$ this yields
$$ \utilde = { 1 \over {2 \mu}} \left[ - \partial_x \ptilde + \rho u_0 \right]
             \left[ {H^2 \over 4} - y^2 \right]
 - {{\rho u_0} \over {3 \mu H^2}} \left[ {H^4 \over {16}}  - y^4 \right] .
$$
If we consider that at the inflow boundary $\utilde=0$, then as the adjoint 
velocity field is also divergence-free, in any section of $x$ we must have:
$$ \int \utilde dy = 0 . $$
This implies:
$$  \int_{-H/2}^{H/2} \utilde \mbox{ dy} = { 1 \over {2 \mu}} \left[ - \partial_x \ptilde + \rho u_0 \right]
             \left[ {H^2 \over 4} y - {y^3 \over 3} \right]_{-H/2}^{H/2}
 - {{\rho u_0} \over {3 \mu H^2}} 
             \left[ {H^4 \over {16}} y  - {y^5 \over 5} \right]_{-H/2}^{H/2}
 = 0 .
$$
Evaluation of all terms leads to the remarkable result:
$$ \partial_x \ptilde = {4 \over 5} \rho u_0 
                = -{{\rho H^2} \over {10 \mu}} \partial_x p \, , $$
i.e. the gradient of the adjoint pressure is also constant and
linearly dependent of $u_0$.
Given that the base level of the
pressure $p$ is arbitrary, we might set it so that it vanishes at the exit, i.e. $p=0$. We finally obtain 
the remarkable result that:
$$ - \ptilde = {{\rho H^2} \over {10 \mu}} p \, , $$
i.e. the pressure and adjoint pressure are related by the factor
${{\rho H^2} \over {10 \mu}}$ and have a constant gradient in
the field. The adjoint velocity is given by:
$$ \utilde = { {\rho u_0} \over {\mu}} \left\{ 
             { 1 \over {10}} \left[ {H^2 \over 4} - y^2 \right]
 - {1 \over {3 H^2}} \left[ {H^4 \over {16}}  - y^4 \right] \right\} \, .
$$
At the center of the channel the velocity is given by:
$$ \utilde(y=0) = {{\rho u_0 H^2} \over {240 \mu}} \, . $$

\bs \noi
\ub{Vortical Energy}:
\ms \noi
If the cost function is given by the vortical energy
$$ I = { 1 \over 2} \int \rho \left| \nabla \times \bm u \right|^2 \mbox{ dx}
                                       \, , $$
then, for the 2-D channel ($u=u(y),v=0,w=z$)
$$ | \nabla \times \bm u |^2 = \left( \partial_y u \right)^2 ~~, $$
so that
$$ I_{u} = \rho \partial_y u (\partial_y u)_{,u} = - \rho \partial_{yy} u 
          = - {\rho \over \mu} \partial_x p 
          = {{8 \rho u_0} \over H^2} \, ,  $$
i.e. constant (!).
Assuming a long channel with no change in $x$ for the variables, the
equation for the adjoint $x$-velocity $\utilde$ simplifies to:
$$ \partial_x \ptilde = \mu \partial_{yy} \utilde - {\rho \over \mu} \partial_x p \, . $$
As this is a long channel and the source-term is constant, the assumption
that $\partial_x \ptilde$ is constant is warranted. This implies that
$\partial_{yy} \utilde$ should also be a constant.
Applying the boundary conditions $\utilde=0$ for $y=-H/2$ and $y=H/2$
yields:
$$ \utilde = \left[ 1 - {4 \over H^2} y^2 \right] \utilde_0 \, . $$
However, if we again consider that at the inflow boundary $\utilde=0$, and given that the adjoint velocity field is divergence-free,
then in any section of $x$ we must have:
$$ \int \utilde dy = 0 \, , $$
which implies that the only possible solution is $\utilde(x,y)=0$, 
and therefore:
$$ -\partial_x \ptilde = {\rho \over \mu} \partial_x p \, . $$
As at the exit the pressure $p$ vanishes, i.e. $p=0$, we finally obtain
the remarkable result that:
$$ -\ptilde = {\rho \over \mu} p \, , $$
i.e. the pressure and adjoint pressure are related by the factor
${\rho \over \mu}$ and have a constant gradient in the field.

\bs \noi
\subsection{Exact Derivatives of Cost Functions}

\bs \noi
\ub{Kinetic Energy}:
$$ I^{ke} = { 1 \over 2 } \int \rho |\bm u|^2 \mbox{ dx} \, .$$
Given that $u=u(y), v=0$ this results in:
$$ I^{ke} = { 1 \over 2 } \rho \int_x dx  \int_y u^2 dy 
          = { 1 \over 2 } \rho L 
            \int u^2_0 \left[ 1 - {4 \over H^2} y^2 \right]^2 dy
$$
$$ I^{ke} = { 1 \over 2 } {{8} \over {15}} L H \rho u^2_0 ~~, $$

$$ I^{ke}_{,u_0} = {{8} \over {15}} L H \rho u_0 
                 = {{2} \over {3}}    H \ptilde_{in} ~~, $$
i.e. linear in the length $L$ and the velocity $u_0$,
and
$$ I^{ke}_{,x} = { 1 \over 2 } {{8} \over {15}} H \rho u^2_0 
               = { 1 \over 2 } {{2} \over {3}}    H \ptilde_{in} u_0 ~~,
$$
i.e. not dependent (constant) of the length $L$ and quadratic in
the velocity $u_0$. In the previous equations we assumed $p_{out}=0$, and used the analytical results that relate mass flow, viscosity and pressure gradient for the Poiseuille flow. 
One should remark that if the domain
that is of interest does not change (e.g. only a certain region
inside the channel is considered), the correct value is:
$$ I^{ke}_{,x} = 0 $$
as the flow is constant in $x$ and therefore the cost functional
does not change if the upstream boundary is moved.

\bs \noi
\ub{Vortical Energy (Dissipation)}:
$$ I^{ve} = { 1 \over 2 } \int \rho | \nabla \times \bm u |^2 \mbox{ dx} \, .$$
Given that $u=u(y), v=0$ this results in: 
$$ I^{ve} = { 1 \over 2 } \rho \int_x dx  \int_y |\partial_y u| ^2 dy 
          = {8 \over 3} {{\rho u^2_0} \over H^2} L H
$$
This implies:
$$ I^{ve}_{,u_0} = {16 \over 3} L \rho {u_0 \over H} 
                 = {{2 L H \ptilde} \over {3}} ~~, $$
i.e. linear in the length $L$ and the velocity $u_0$,
and
$$ I^{ve}_{,x} = {8 \over 3} {{\rho u^2_0} \over H} 
               = {{L H \ptilde u_0} \over {3}} ~~, $$
i.e. not dependent (constant) of the length $L$ and quadratic in
the velocity $u_0$. Notice, though, that as before if the domain
that is of interest does not change (e.g. only a certain region
inside the channel is considered), the correct value is:
$$ I^{ve}_{,x} = 0 $$
as the flow is constant in $x$ and the cost functional will not change
if the upstream boundary is moved.
%

\bibliographystyle{abbrv}

\bibliography{cas-refs}

\end{document}

%% file: abbrevs.tex
\def\ms{\medskip}
\def\bs{\bigskip}
\def\ub{\underbar}
\def\noi{\noindent}

\def\lvec{{\bf l}}

\def\ptilde{\tilde{p}}
\def\utilde{\tilde{u}}
\def\vtilde{\tilde{v}}

\def\Rvec{{\bf R}}

\def\Amat{{\bf A}}
\def\Bmat{{\bf B}}

\def\pmb#1{\setbox0=\hbox{$#1$}%
             \kern-.027em\copy0\kern-\wd0
             \kern+.009em\copy0\kern-\wd0
             \kern+.009em\copy0\kern-\wd0
             \kern+.009em\copy0\kern-\wd0
             \kern+.009em\copy0\kern-\wd0
             \kern+.009em\copy0\kern-\wd0
             \kern+.009em\copy0\kern-\wd0
             \kern-.045em\raise+.012em\copy0\kern-\wd0
             \kern+.009em\raise+.012em\copy0\kern-\wd0
             \kern+.009em\raise+.012em\copy0\kern-\wd0
             \kern+.009em\raise-.012em\copy0\kern-\wd0
             \kern+.009em\raise-.012em\copy0\kern-\wd0
             \kern-.018em\copy0\kern-\wd0\raise-.012em\box0}
\def\Pmb#1{\setbox0=\hbox{$#1$}%
             \kern-.033em\copy0\kern-\wd0
             \kern+.011em\copy0\kern-\wd0
             \kern+.011em\copy0\kern-\wd0
             \kern+.011em\copy0\kern-\wd0
             \kern+.011em\copy0\kern-\wd0
             \kern+.011em\copy0\kern-\wd0
             \kern+.011em\copy0\kern-\wd0
             \kern-.055em\raise+.015em\copy0\kern-\wd0
             \kern+.011em\raise+.015em\copy0\kern-\wd0
             \kern+.011em\raise+.015em\copy0\kern-\wd0
             \kern+.011em\raise-.015em\copy0\kern-\wd0
             \kern+.011em\raise-.015em\copy0\kern-\wd0
             \kern-.022em\copy0\kern-\wd0\raise-.015em\box0}
\def\Grd{\nabla}

\vfill